\documentclass[aps,preprint,showpacs,groupedaddress]{revtex4-1}
\usepackage{amssymb}
\usepackage{mathrsfs}
\usepackage{amsmath}
\usepackage{mathtools}
\usepackage{pstricks}
\usepackage{color}
\usepackage{graphicx}
\usepackage{amsthm}

\usepackage{physics}

\begin{document}


\title{\bf Universality hypothesis breakdown at one-loop order}



\author{P. R. S. Carvalho}
\email{prscarvalho@ufpi.edu.br}
\affiliation{\it Departamento de F\'\i sica, Universidade Federal do Piau\'\i, 64049-550, Teresina, PI, Brazil}





\begin{abstract}
We probe the universality hypothesis by analytically computing, at least, the two-loop corrections to the critical exponents for $q$-deformed O($N$) self-interacting $\lambda\phi^{4}$ scalar field theories through six distinct and independent field-theoretic renormalization group methods and $\epsilon$-expansion techniques. We show that the effect of $q$-deformation on the one-loop corrections to the $q$-deformed critical exponents is null, so universality hypothesis is broken down at this loop order. Such effect emerges just, at least, at two-loop level and the validity of universality hypothesis is restored. The $q$-deformed critical exponents obtained through the six methods are the same and, furthermore, reduce to their non-deformed values in the appropriated limit. 
\end{abstract}


\maketitle


\section{Introduction} 

\par The critical behavior of completely different systems as a fluid and a ferromagnet can be described at the same footing, since they present an identical set of critical exponents. In fact, when this happens, we say that the distinct systems belong to the same universality class. An universality class is characterized by many different systems with the same set of critical exponents. The critical exponents depend jointly on the dimension $d$, $N$ and symmetry of some $N$-component order parameter of the systems if the interactions are of short- or long-rage type. Otherwise, they do not depend on the details of the systems as the form of the interactions and their critical temperature. The order parameter is responsible for revealing the presence of an ordered phase in the system which is related to a broken symmetry. In the other phase, in the disordered one, the symmetry is intact. The properties of both ordered and disordered phases can be encompassed, in the field-theoretic formulation of phase transitions and critical phenomena, by defining a fluctuating quantum field whose mean value can be identified to the order parameter. In the ordered phase its mean value is non-vanishing, while in the disordered one it is null. Thus, the parameters characterizing an universality class are intimately related to the general properties of the field as its symmetry and number $N$ of components, besides the dimension of space-time where it is embedded. Any change in the general properties of the field must be accompanied by changes in the values of the critical exponents. This is the content of universality hypothesis. The values of the critical exponents are a result of the non-trivial interactions among the many degrees of freedom at various length scales. This is the essence of renormalization group tool introduced by Wilson \cite{Wilson197475}. These ideas, proportioned the computation of corrections, in the dimensional parameter $\epsilon = 4 - d$, to Landau theory. In the Landau theory, the interactions at many length scales are neglected and the corresponding critical exponents, the Landau ones are obtained \cite{Stanley}. The corrections to Landau critical exponents can be obtained and now the interactions among the many length scales are taken into account and represented by the loop radiative quantum corrections to the exponents. Thus, how much sensible is a physical effect, more and more loops must be computed for describing precisely that effect. These quantum corrections are plagued by divergences and have to be ruled out by some mathematical procedure. Technically, these divergences originate in the interactions among the many values of the quantum field at the same point of space-time. Specifically, they originate from the commutation relations between the destruction and creation operators representing the quantum field. In this paper, we examine the effect of modifications of the commutation relations of the quantum field on the values of the critical exponents. For that, we investigate the modified properties of the so called $q$-deformed O($N$) self-interacting $\lambda\phi^{4}$ scalar field theory \cite{G.Vinod}. The $q$-deformation idea has motivated applications in many research areas as Relativistic fermion scattering \cite{Adv.High.EnergyPhys.20179530874}, Ramsauer-Townsend effect \cite{Eur.Phys.J.Plus1322017398}, Boson algebra related to gentile statistics \cite{Int.J.Theor.Phys.5620171746}, Berry phase \cite{EPL11320162000}, dark matter and dark energy \cite{Phys.DarkUniv.1620171}, Dirac oscillator \cite{Adv.HighEnergyPhys.20179371391}, Yang-Mills theory \cite{JHEP1220160630} and cat states \cite{Phys.Rev.D912015044024} for citing just a few examples. The $q$-deformed scalar field in free space-time is given by
\begin{eqnarray}\label{ygfdxzsze}
\phi_{q}(x) = \int \frac{d^{3}k}{(2\pi)^{3/2}2\omega_{k}^{1/2}}[a(k)_{q}\exp^{-ikx} +a_{q}^{\dagger}(k)\exp^{ikx}]\quad
\end{eqnarray}
where $\omega_{k}^{2} = \vec{k}^{2} + m^{2}$ and its creation and destruction operators obey to the $q$-deformed commutation relations
\begin{eqnarray}\label{kljkkjij}
[a(k)_{q}, a_{q}^{\dagger}(k^{\prime})] = q^{N(k)}\delta(k - k^{\prime}), 
\end{eqnarray}  
\begin{eqnarray}\label{sfcvcgf}
[a(k)_{q}, a_{q}(k^{\prime})] = 0 = [a(k)_{q}^{\dagger}, a_{q}^{\dagger}(k^{\prime})], 
\end{eqnarray}    
where $N(k) = a_{q}^{\dagger}(k)a_{q}(k)$. The corresponding general massive free $q$-propagator in momentum space-time is given by $G_{0}(P) = \parbox{12mm}{\includegraphics[scale=1.0]{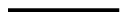}} = q/(P^{2} + m^{2})$. In this work we have to show that the $q$-deformation parameter $q$ leads to non-trivial next-to-leading level critical exponents. All divergences of the theory to be renormalized are contained in the initially divergent correlation functions or primitively divergent $1$PI vertex parts $\Gamma^{(2)}$ and $\Gamma^{(4)}$. If we renormalize them, all higher $1$PI vertex parts are automatically renormalized, since they are composed of the primitive ones through a skeleton expansion \cite{ZinnJustin}. We observe in eq. (\ref{kljkkjij}) that in the limit $q \rightarrow 1$, we recover the non-deformed quantum field properties. As the critical exponents are universal quantities, they can be obtained in theories representing the system at and near the critical point and must be the same when obtained through the different theories. In the field-theoretic formulation of the problem approached in this paper, the system at (near) the critical point is described by an infrared divergent massless (ultraviolet divergent massive) theory at dimensions less than four, \emph{i. e.} $d < 4$ (for $d = 4$, we have a Gaussian theory and the critical exponents are the Landau ones and the range $d \geq 4$ leads also to Landau critical exponents \cite{Stanley}), valid for $2 < d < 4$, since the mass in this theoretic formulation plays the role of the difference between an arbitrary temperature and the critical one $T - T_{c}$, thus when $T = T_{c}$, $m^{2} = 0$. This fact and the one that a given massless and massive theory can be renormalized at distinct renormalization group schemes, permit us to apply six versions of descriptions of the system, a massless (critical) theory renormalized at three different and independent renormalization schemes and similarly three massive (non-critical) ones for employing the referred task. Thus, we have the advantage of computing the critical exponents through the many different methods and checking the final results. We have to obtain the same values for the critical exponents because they are universal quantities, although the corresponding $q$-deformed $\beta_{q}$-function, anomalous dimensions and fixed points present distinct values in the different methods.

\par In this work, we evaluate analytically, at least at next-to-leading loop level, the critical exponents for $q$-deformed O($N$) self-interacting $\lambda\phi^{4}$ scalar field theories for probing the universality hypothesis. For that, we employ six different and independent field-theoretic renormalization group methods based on dimensional regularization and $\epsilon$-expansion techniques. The three first methods are applied for a critical theory and the three last ones for a non-critical one. We present the results for the $q$-deformed critical exponents and give both mathematical and physical interpretation for them. At the end, we present our conclusions.   

\section{At the critical point}\label{At the critical point}

\par In the critical theory, we can obtain the critical exponents through three independent methods to be displayed below.

\subsection{Normalization conditions}\label{Normalization conditions} We begin our journey of computing the critical exponents by applying the normalization conditions method \cite{BrezinLeGuillouZinnJustin,Amit}. In this method, we start from the bare massless theory with the Feynman diagrams $\parbox{10mm}{\includegraphics[scale=.8]{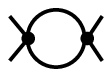}}$, $\parbox{10mm}{\includegraphics[scale=.6]{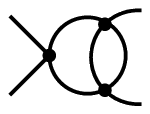}}$, $\parbox{14mm}{\includegraphics[scale=.8]{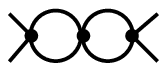}}$, $\parbox{10mm}{\includegraphics[scale=.8]{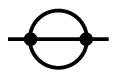}}$, $\parbox{8mm}{\includegraphics[scale=.6]{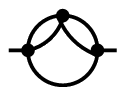}}$, $\parbox{12mm}{\includegraphics[scale=.8]{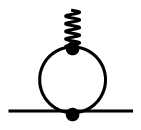}}$, $\parbox{12mm}{\includegraphics[scale=.6]{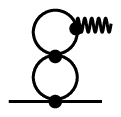}}$ and $\parbox{10mm}{\includegraphics[scale=.6]{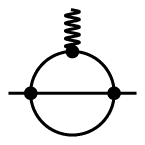}}$. From all these diagrams, just a minimal set of four of them are needed to be evaluated. The external momenta of the needed ones $\parbox{10mm}{\includegraphics[scale=.8]{fig10.eps}}$, $\parbox{10mm}{\includegraphics[scale=.8]{fig6.eps}}$, $\parbox{10mm}{\includegraphics[scale=.6]{fig7.eps}}$ and $\parbox{10mm}{\includegraphics[scale=.6]{fig21.eps}}$ are held at fixed values through normalization conditions which define the symmetry point $P^{\prime 2} = 1$ to be employed. Then, we have to compute $\parbox{6mm}{\includegraphics[scale=0.6]{fig10.eps}}_{SP} \equiv \parbox{6mm}{\includegraphics[scale=0.6]{fig10.eps}}\vert_{P^{\prime 2}=1}$, $\parbox{8mm}{\includegraphics[scale=0.7]{fig6.eps}}^{\prime} \equiv (\partial \parbox{8mm}{\includegraphics[scale=0.7]{fig6.eps}}/\partial P^{\prime 2})\vert_{P^{\prime 2}=1}$, $\parbox{8mm}{\includegraphics[scale=0.5]{fig21.eps}}_{SP} \equiv \parbox{8mm}{\includegraphics[scale=0.5]{fig21.eps}}\vert_{P^{\prime 2}=1}$ and $\parbox{8mm}{\includegraphics[scale=0.6]{fig7.eps}}^{\prime} \equiv (\partial \parbox{8mm}{\includegraphics[scale=0.6]{fig7.eps}}/\partial P^{\prime 2})\vert_{P^{\prime 2}=1}$, where $P^{\prime}$ is written in terms of some momentum scale unit $\kappa$. Thus the $q$-deformed $\beta_{q}$-function and anomalous dimensions are given by
\begin{eqnarray}\label{jkjhigyg}
\beta_{q}(u) = -\epsilon u +   \frac{N + 8}{6}\left( 1 + \frac{1}{2}\epsilon \right)q^{2}u^{2} -  \frac{3N + 14}{12}q^{4}u^{3} + \frac{N + 2}{36}q^{3}(1-q)u^{3}, 
\end{eqnarray}
\begin{eqnarray}
\gamma_{\phi ,q} = \frac{N + 2}{72}\left( 1 + \frac{5}{4}\epsilon \right)q^{3}u^{2} -  \frac{(N + 2)(N + 8)}{864}q^{5}u^{3},  
\end{eqnarray}
\begin{eqnarray}\label{sfsvcfdt}
\overline{\gamma}_{\phi^{2}, q}(u) = \frac{N + 2}{6}\left( 1 + \frac{1}{2}\epsilon \right)q^{2} u -  \frac{N + 2}{12}q^{4}u^{2},
\end{eqnarray}
where $\overline{\gamma}_{\phi^{2}}(u) = \gamma_{\phi^{2}}(u) - \gamma_{\phi}(u)$. We observe that in this renormalization scheme, the $q$-deformed $\beta_{q}$-function and anomalous dimensions are finite functions as required by any renormalization program and depend on the symmetry point employed through their second, first and first terms, respectively. We will show later that the $q$-deformed critical exponents do not depend on this non-universal feature.

\subsection{Minimal subtraction scheme} In the minimal subtractions scheme \cite{BrezinLeGuillouZinnJustin,Amit}, once again we start from the bare massless theory and the external momenta of the minimal set needed Feynman diagrams above now are left at arbitrary values, which shows its generality and elegance since the $q$-deformed $\beta_{q}$-function and anomalous dimensions do not depend on specific values of the external momenta and thus do not depend on any symmetry point values. The corresponding $\beta_{q}$-function and anomalous dimensions have the form  
\begin{eqnarray}\label{rygyfggffgyt}
\beta_{q}(u) = -\epsilon u +   \frac{N + 8}{6}q^{2}u^{2} -  \frac{3N + 14}{12}q^{4}u^{3} + \frac{N + 2}{36}q^{3}(1-q)u^{3}, 
\end{eqnarray}
\begin{eqnarray}\label{hkfusdrs}
\gamma_{\phi ,q} = \frac{N + 2}{72}q^{3}u^{2} - \frac{(N + 2)(N + 8)}{1728}q^{5}u^{3},  
\end{eqnarray}
\begin{eqnarray}\label{kujyhghsghju}
\overline{\gamma}_{\phi^{2}, q}(u) = \frac{N + 2}{6}q^{2} u -  \frac{N + 2}{12}q^{4}u^{2}.
\end{eqnarray}

\subsection{Massless Bogoliubov-Parasyuk-Hepp-Zimmermann method} 

\par In the massless Bogoliubov-Parasyuk-Hepp-Zimmermann (BPHZ) method [13,14], as opposed to the last ones, the divergences are removed of the massless theory through the introduction of counterterms to the loop expansions for the $1$PI vertex parts at a given loop level. We repeat this procedure order by order in perturbation theory for attaining the renormalized theory. Besides the initially minimal set of diagrams of Normalization conditions and Minimal subtraction scheme methods, we have additionally to compute the counterterms $\parbox{10mm}{\includegraphics[scale=.9]{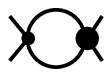}}$, $\parbox{10mm}{\includegraphics[scale=.9]{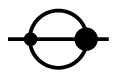}}$, $\parbox{8mm}{\includegraphics[scale=.15]{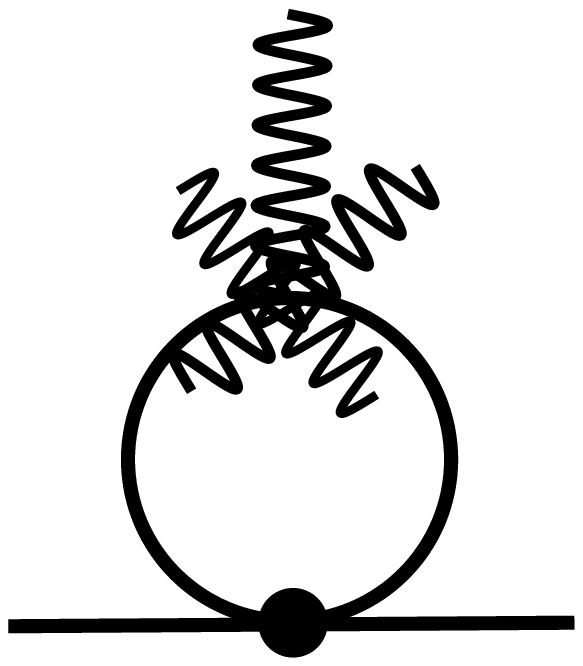}}$ and $\parbox{8mm}{\includegraphics[scale=.15]{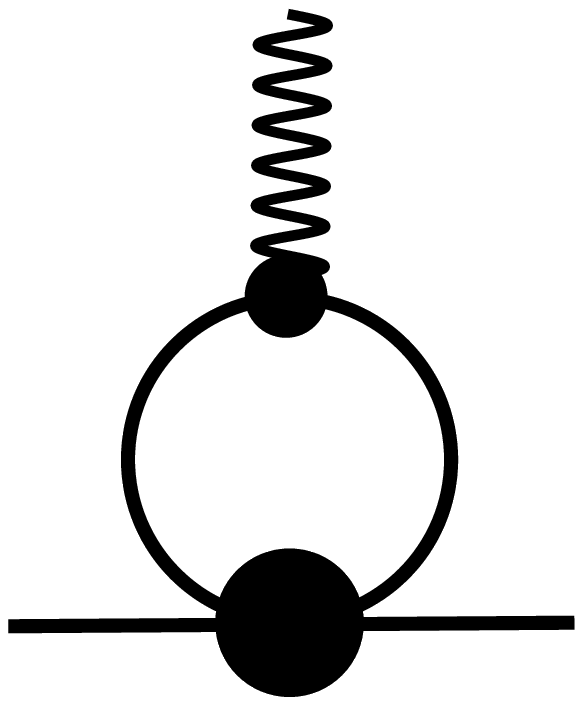}}$, where
\begin{eqnarray}
&&\parbox{10mm}{\includegraphics[scale=1.0]{fig25.eps}}\equiv \parbox{12mm}{\includegraphics[scale=1.0]{fig10.eps}}\Bigg|_{-\mu^{\epsilon}u\rightarrow -\mu^{\epsilon}uc_{u}^{1}}, 
\end{eqnarray}
\begin{eqnarray}
&&\parbox{12mm}{\includegraphics[scale=1.0]{fig26.eps}}\equiv \parbox{12mm}{\includegraphics[scale=1.0]{fig6.eps}}\Bigg|_{-\mu^{\epsilon}u\rightarrow -\mu^{\epsilon}uc_{u}^{1}},
\end{eqnarray}
\begin{eqnarray}
&&\parbox{10mm}{\includegraphics[scale=.2]{fig31.eps}}\quad\equiv \parbox{14mm}{\includegraphics[scale=1.0]{fig14.eps}}\Bigg|_{-u\rightarrow -uc_{\phi^{2}}^{1}},
\end{eqnarray}
\begin{eqnarray}
&&\parbox{10mm}{\includegraphics[scale=.2]{fig32.eps}}\quad\equiv \parbox{14mm}{\includegraphics[scale=1.0]{fig14.eps}}\Bigg|_{-u\rightarrow -uc_{u}^{1}}
\end{eqnarray}
and $c_{u}^{1}$ and $c_{\phi^{2}}^{1}$ are the counterterms at one-loop order given by
\begin{eqnarray}
&&\parbox{6mm}{\includegraphics[scale=.1]{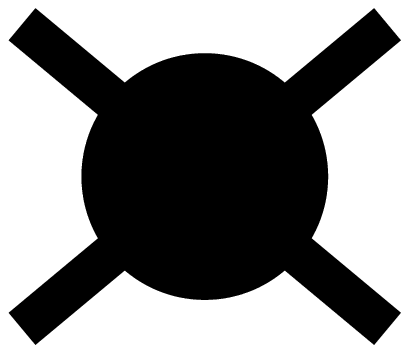}} = -\mu^{\epsilon}uc_{u}^{1} = -\frac{3}{2} \mathcal{K} 
\left(\parbox{10mm}{\includegraphics[scale=1.0]{fig10.eps}} \right),
\end{eqnarray}
\begin{eqnarray}
&&\parbox{16mm}{\includegraphics[scale=.2]{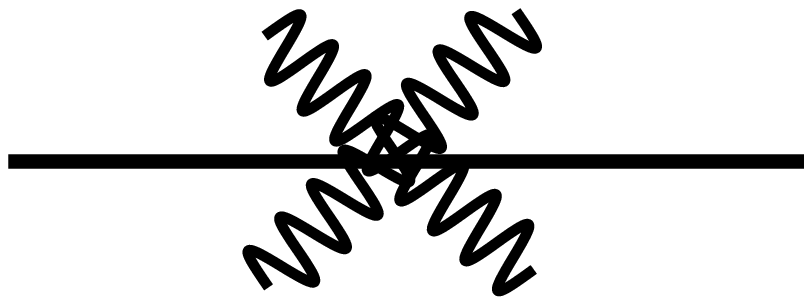}} = -c_{\phi^{2}}^{1} = -\frac{1}{2} \mathcal{K} 
\left(\parbox{11mm}{\includegraphics[scale=.8]{fig14.eps}} \right).
\end{eqnarray}
These counterterms are not independent and we can show one more time that the only independent diagrams to be computed are the ones of the last method. Thus, the $q$-deformed $\beta_{q}$-function and the field anomalous dimension are the same as the ones in eqs. (\ref{rygyfggffgyt}) and (\ref{hkfusdrs}), respectively. The composite field anomalous dimension has the form  
\begin{eqnarray}\label{srdsrdsr}
\gamma_{\phi^{2}, q}(u) = \frac{N + 2}{6}q^{2} u -  \frac{5(N + 2)}{72}q^{4}u^{2} +  \frac{N + 2}{72}q^{3}(1-q)u^{2},
\end{eqnarray}
where in this method we compute $\gamma_{\phi^{2}, q}(u)$ instead $\overline{\gamma}_{\phi^{2}, q}(u)$ as in the earlier method.

\section{Near the critical point} 

\par In the non-critical situation, we can evaluate the $q$-deformed critical exponents by applying another three distinct and independent methods as well.

\subsection{Callan-Symanzik method} This method \cite{BrezinLeGuillouZinnJustin,Amit} treats a massive theory. A massive theory is more general than the massless one studied in the last Sec., since the diagrams with tadpole insertions $\parbox{10mm}{\includegraphics[scale=.9]{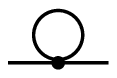}}$, $\parbox{10mm}{\includegraphics[scale=.9]{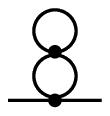}}$, $\parbox{8mm}{\includegraphics[scale=.9]{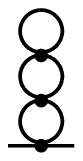}}$, $\parbox{14mm}{\includegraphics[scale=.9]{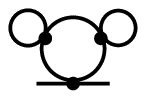}}$, $\parbox{12mm}{\includegraphics[scale=.8]{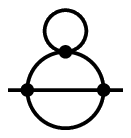}}$, $\parbox{10mm}{\includegraphics[scale=.9]{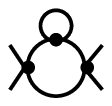}}$, $\parbox{14mm}{\includegraphics[scale=.8]{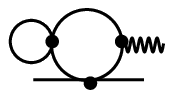}}$ and the one which is independent of external momenta $\parbox{8mm}{\includegraphics[scale=.9]{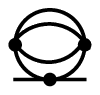}}$ are null in the massless theory and nonvanishing in the massive one and now must be evaluated. Through a mathematical trick, these additional diagrams are eliminated by redefining the initial bare mass at tree-level, thus being substituted to its three-loop counterpart. We then end up with an effective loop expansion for the diagrams with terms like $\left(\parbox{12mm}{\includegraphics[scale=.8]{fig6.eps}}  - \parbox{12mm}{\includegraphics[scale=.8]{fig6.eps}}\bigg|_{P^{\prime 2}=0}\right)$ and $\left(\parbox{12mm}{\includegraphics[scale=.7]{fig7.eps}}  - \parbox{12mm}{\includegraphics[scale=.7]{fig7.eps}}\bigg|_{P^{\prime 2}=0}\right)$. As the diagrams $ \parbox{12mm}{\includegraphics[scale=.8]{fig6.eps}}\bigg|_{P^{\prime 2}=0}$ and $\parbox{12mm}{\includegraphics[scale=.7]{fig7.eps}}\bigg|_{P^{\prime 2}=0}$ do not depend on external momenta, they do not contribute to the calculation through the derivative with respect to $P^{\prime 2}$ defining the symmetry point in this method displayed below. Thus, we have to evaluate only a minimal set of four of them. As the bare theory is a massive one, the external momenta of the minimal set of needed Feynman diagrams can be held at fixed vanishing values and the symmetry point is now $P^{\prime 2} = 0$. The minimal set of needed diagrams are the massive versions of the ones computed in the Normalization conditions method, namely 
$\parbox{6mm}{\includegraphics[scale=0.6]{fig10.eps}}_{SP} \equiv \parbox{6mm}{\includegraphics[scale=0.6]{fig10.eps}}\vert_{P^{\prime 2}=0}$, $\parbox{8mm}{\includegraphics[scale=0.7]{fig6.eps}}^{\prime} \equiv (\partial \parbox{8mm}{\includegraphics[scale=0.7]{fig6.eps}}/\partial P^{\prime 2})\vert_{P^{\prime 2}=0}$, $\parbox{8mm}{\includegraphics[scale=0.5]{fig21.eps}}_{SP} \equiv \parbox{8mm}{\includegraphics[scale=0.5]{fig21.eps}}\vert_{P^{\prime 2}=0}$ and $\parbox{8mm}{\includegraphics[scale=0.6]{fig7.eps}}^{\prime} \equiv (\partial \parbox{8mm}{\includegraphics[scale=0.6]{fig7.eps}}/\partial P^{\prime 2})\vert_{P^{\prime 2}=0}$
Thus the $q$-deformed $\beta_{q}$-function and anomalous dimensions are given by
\begin{eqnarray}\label{fhdfghg}
\beta_{q}(u) = -\epsilon u +   \frac{N + 8}{6}\left( 1 - \frac{1}{2}\epsilon \right)q^{2}u^{2} - \frac{3N + 14}{12}q^{4}u^{3} + \frac{N + 2}{36}q^{3}(1-q)u^{3}, 
\end{eqnarray}
\begin{eqnarray}
\gamma_{\phi ,q} = \frac{N + 2}{72}\left( 1 - \frac{1}{4}\epsilon + I\epsilon \right)q^{3}u^{2} -  \frac{(N + 2)(N + 8)}{432}\left( 1 + I \right)q^{5}u^{3},  
\end{eqnarray}
\begin{eqnarray}\label{sfdtsvdb}
\overline{\gamma}_{\phi^{2}, q}(u) = \frac{N + 2}{6}\left( 1 - \frac{1}{2}\epsilon \right)q^{2} u -  \frac{N + 2}{12}q^{4}u^{2},
\end{eqnarray}
where the integral $I$ \cite{Amit,Phys.Rev.D841973434,Carvalho2009178,Carvalho2010151} is a number and a residual consequence of the symmetry point chosen. The integral $I$ can be calculated analytically in terms of dilogarithm function of certain argument \cite{Ramond}. At least up to the loop level considered in this work, $I$ is canceled out in the $q$-deformed critical exponents computation. One more time, the $q$-deformed $\beta_{q}$-function and anomalous dimensions depend of the symmetry point employed through their second, first and first terms, respectively.

\subsection{Unconventional Minimal subtraction scheme} This method was introduced recently by one of the authors \cite{J.Math.Phys.542013093301} and was inspired in the last one but now for arbitrary values of external momenta.  Now, the price to pay for maintaining the external momenta at general values, the diagrams $ \parbox{12mm}{\includegraphics[scale=.8]{fig6.eps}}\bigg|_{P^{\prime 2}=0}$ and $\parbox{12mm}{\includegraphics[scale=.7]{fig7.eps}}\bigg|_{P^{\prime 2}=0}$ do not disappear of the intermediate results though derivatives with respect to $P^{\prime 2}$ and we have to compute them at the next order in the dimensional expansion parameter $\epsilon$. At the end of calculations they do not contribute to the corresponding $q$-deformed $\beta_{q}$-function and anomalous dimensions, which have the same form as the corresponding ones of Minimal subtraction scheme (\ref{rygyfggffgyt})-(\ref{kujyhghsghju}).

\subsection{Massive Bogoliubov-Parasyuk-Hepp-Zimmermann method} This method \cite{BogoliubovParasyuk,Hepp,Zimmermann,Kleinert} is the most general one of all six treated here, since it is dealing with a massive theory at arbitrary values of external momenta and with its initial bare mass at its tree-level value. Thus, we have that, necessarily, to compute diagrams like the $\parbox{9mm}{\includegraphics[scale=.8]{fig1.eps}}$ and $\parbox{8mm}{\includegraphics[scale=.7]{fig2.eps}}$ ones for example. The corresponding $q$-deformed $\beta_{q}$-function and anomalous dimensions have the same form as that obtained in the massless version of this method, \emph{i.e.} the Massless Bogoliubov-Parasyuk-Hepp-Zimmermann method.

\section{Results for $q$-deformed critical exponents and discussion}

\par Now, we can compute the loop corrections to $q$-deformed critical exponents in the six methods. For that, we employ the relations $\eta\equiv\gamma_{\phi,q}(u^{*})$ and $\nu_{q}^{-1}\equiv 2 - \eta_{q} - \overline{\gamma}_{\phi^{2},q}(u^{*})$ in the first, second, fourth and fifth methods and the $\eta_{q}\equiv\gamma_{\phi,q}(u^{*})$ and $\nu_{q}^{-1}\equiv 2 - \gamma_{\phi^{2},q}(u^{*})$ ones in the third and last renormalization schemes for evaluating, independently, $\eta_{q}$ and $\nu_{q}$, since there are six $q$-deformed critical exponents to be computed and four scaling relations among them \cite{Stanley}. The $u^{*}$ quantity is the non-trivial fixed point which gives the fluctuation corrections to mean field $q$-deformed critical exponents. For a given method, it is obtained as the non-trivial solution to the equation $\beta_{q}(u^{*}) = 0$ for the corresponding $q$-deformed $\beta_{q}$-function of the referred method. The trivial or Gaussian solution $u^{*} = 0$ gives no loop corrections to the $q$-deformed critical exponents, thus leading to their trivial mean field values. We have computed the $q$-deformed critical exponents, at least, up to two-loop order and obtained the same values in the six methods. They are
\begin{eqnarray}
\alpha_{q} = \frac{(4 - N)}{4(N + 8)}\epsilon +  \frac{(N + 2)(N^{2} + 30N + 56)}{4(N + 8)^{3}}\epsilon^{2} -  \frac{(N + 2)(4 - N)}{2(N + 8)^{3}}\frac{(1 - q)}{q}\epsilon^{2},
\end{eqnarray}
\begin{eqnarray}
\beta_{q} = \frac{1}{2} - \frac{3}{2(N + 8)}\epsilon +  \frac{(N + 2)(2N + 1)}{2(N + 8)^{3}}\epsilon^{2} +  \frac{3(N + 2)}{2(N + 8)^{3}}\frac{(1 - q)}{q}\epsilon^{2},
\end{eqnarray}
\begin{eqnarray}
\gamma_{q} = 1 + \frac{(N + 2)}{2(N + 8)}\epsilon +  \frac{(N + 2)(N^{2} + 22N + 52)}{4(N + 8)^{3}}\epsilon^{2} -  \frac{(N + 2)^{2}}{2(N + 8)^{3}}\frac{(1 - q)}{q}\epsilon^{2},
\end{eqnarray}
\begin{eqnarray}
\delta_{q} = 3 +  \epsilon +  \frac{N^{2} + 14N + 60}{2(N + 8)^{2}}\epsilon^{2} - \frac{N + 2}{(N + 8)^{2}}\frac{(1 - q)}{q}\epsilon^{2},
\end{eqnarray}
\begin{eqnarray}
\nu_{q} = \frac{1}{2} + \frac{(N + 2)}{4(N + 8)}\epsilon +  \frac{(N + 2)(N^{2} + 23N + 60)}{8(N + 8)^{3}}\epsilon^{2} + \frac{(N + 2)(4 - N)}{8(N + 8)^{3}}\frac{(1 - q)}{q}\epsilon^{2},
\end{eqnarray}
\begin{eqnarray}
\eta_{q} = \frac{(N + 2)}{2(N + 8)^{2}q}\epsilon^{2} \left\{ 1 + \left[ \frac{6(3N + 14)}{(N + 8)^{2}} -\frac{1}{4} - \frac{2(N + 2)}{(N + 8)^{2}}\frac{(1 - q)}{q} \right]\epsilon\right\}.\quad\quad
\end{eqnarray}
First of all, as we have written the $q$-deformed critical exponents such that it is an easy task to take the limit $q \rightarrow 1$, we can easily see that they reduce to their corresponding non-deformed values \cite{Amit} in that limit, as expected. Secondly, we realize that all one-loop corrections to $q$-deformed critical exponents are the same as their non-deformed counterparts. This means that the universality hypothesis was broken down at that loop order, \emph{i.e.} that a change in the internal properties of the fluctuating field did not affect the universal $q$-deformed critical exponents. We can furnish a mathematical explanation for this result by extending a known fact, valid particularly for Lorentz-violating systems for example, that a possible new physical effect represented by a given parameter can not turn out to be a real physical effect if this parameter can be eliminated from the Lagrangian density through coordinate redefinitions \cite{Phys.Rev.D832011016013}. In that case, the critical exponents for a similar non-deformed scalar theory, now with Lorentz violation \cite{PhysRevD.84.065030,Carvalho2013850,Carvalho2014320}, did not present any modifications with respect to their Lorentz-invariant counterparts \cite{EurophysLett.108.21001,Int.J.Mod.Phys.B.30.1550259,Int.J.Geom.MethodsMod.Phys.13.1650049}. In our case, if we restrict our analysis to the one-loop level, we observe that the $q$-deformed results can be obtained from the non-deformed ones by a simple rescaling of $q$-deformed coupling constant in terms of its non-deformed counterpart as $u = q u^{(0)}$, where $u^{(0)}$ is the non-deformed dimensionless renormalized coupling constant. This can be understood if we remember that for a general diagram, the number of loops $l$, internal lines $i$ and order $n$ of interactions are related by $l = i - n +1$ \cite{Ryder}. The only way we have $i = n$, the internal lines number being equal to the interaction order number, is for $l = 1$ (one-loop level) and we can state the rescaling above, since each internal line is accompanied of a power of $q$, then the $i = n$ equality implies that, for one-loop level, $q$ and $u$ are proportional. If we try to find any possible rescaling for higher loop orders, we will arrive at the conclusion that such rescaling is impossible, then the $q$-deformed theory at higher loops is non-trivial and can not be obtained from its non-deformed counterpart through a simple rescaling, thus being necessary to compute the corresponding loop quantum corrections to the $q$-deformed critical exponents for verifying the restoration of universality hypothesis validity. The physical interpretation for this result for one-loop level is that the $q$-deformation is a so slight modification of the internal properties of the fluctuating field, via commutation relations of its creation and destruction operators, that this modification is exhibited in the loop corrections to the $q$-deformed critical exponents just, as least, at two-loop order as we can observe in the two-loop results for the $q$-deformed critical exponents (and three-loop term of $\eta_{q}$). Thus, we have verified the universality hypothesis restoration. We can conjecture that the $q$-deformation manifests at all loop orders as argued above as well and thus that the universality hypothesis is valid for all loop levels, since the last term of the $\eta_{q}$ critical index is of three-loop level and presents a $q$-deformed three-loop correction. Now we present our conclusions.

\section{Conclusions}

\par We computed analytically the critical exponents for $q$-deformed O($N$) self-interacting $\lambda\phi^{4}$ scalar field theories, at least, at two-loop order for probing the universality hypothesis. For that, we employed six distinct and independent field-theoretic renormalization group methods and $\epsilon$-expansion techniques. We showed that the one-loop corrections to the $q$-deformed critical exponents were not affected by the $q$-deformation mechanism thus showing the breaking down of the universality hypothesis at that loop level. We presented the mathematical explanation for this result: a possible new physical effect represented by a given parameter can occur, in fact, if this parameter can not be eliminated from the Lagrangian density by coordinate redefinitions. We showed that at one-loop level, such coordinate redefinition can be made. The corresponding physical interpretation is that the $q$-deformation mechanism is a so slight one that it does not manifest at the lower loop level, the one-loop one. The $q$-deformation mechanism can be perceived just, at least, at two-loop order. That was the case as shown, at least, in the two-loop corrections to the $q$-deformed critical exponents, thus restoring the universality hypothesis validity. We also presented a mathematical explanation, through similar arguments for the one-loop order case, that, on the other hand, for higher loop levels, we can not redefine coordinates such that the $q$-deformed theory can be obtained from the corresponding non-deformed one. Thus, the higher-loop corrections to the $q$-deformed critical exponents can be obtained only by their explicit computations. This implies that we can conjecture the universality hypothesis validity for all loop orders. This conjecture is confirmed, at least, at three-loop level through the computed last and three-loop order $q$-deformed critical index $\eta_{q}$ value. The dependence of $q$-deformed critical exponents on the $q$-deformation parameter $q$ opens the possibility of detecting its influence, through future experiments, on measured $q$-deformed critical exponents.

\section*{Acknowledgements}

\par  PRSC would like to thank the kind referee for helpful comments and Federal University of Piau\'{i} for financial support.

\bibliography{apstemplate}

\providecommand{\noopsort}[1]{}\providecommand{\singleletter}[1]{#1}%
\begin{thebibliography}{31}%
\makeatletter
\providecommand \@ifxundefined [1]{%
 \@ifx{#1\undefined}
}%
\providecommand \@ifnum [1]{%
 \ifnum #1\expandafter \@firstoftwo
 \else \expandafter \@secondoftwo
 \fi
}%
\providecommand \@ifx [1]{%
 \ifx #1\expandafter \@firstoftwo
 \else \expandafter \@secondoftwo
 \fi
}%
\providecommand \natexlab [1]{#1}%
\providecommand \enquote  [1]{``#1''}%
\providecommand \bibnamefont  [1]{#1}%
\providecommand \bibfnamefont [1]{#1}%
\providecommand \citenamefont [1]{#1}%
\providecommand \href@noop [0]{\@secondoftwo}%
\providecommand \href [0]{\begingroup \@sanitize@url \@href}%
\providecommand \@href[1]{\@@startlink{#1}\@@href}%
\providecommand \@@href[1]{\endgroup#1\@@endlink}%
\providecommand \@sanitize@url [0]{\catcode `\\12\catcode `\$12\catcode
  `\&12\catcode `\#12\catcode `\^12\catcode `\_12\catcode `\%12\relax}%
\providecommand \@@startlink[1]{}%
\providecommand \@@endlink[0]{}%
\providecommand \url  [0]{\begingroup\@sanitize@url \@url }%
\providecommand \@url [1]{\endgroup\@href {#1}{\urlprefix }}%
\providecommand \urlprefix  [0]{URL }%
\providecommand \Eprint [0]{\href }%
\providecommand \doibase [0]{http://dx.doi.org/}%
\providecommand \selectlanguage [0]{\@gobble}%
\providecommand \bibinfo  [0]{\@secondoftwo}%
\providecommand \bibfield  [0]{\@secondoftwo}%
\providecommand \translation [1]{[#1]}%
\providecommand \BibitemOpen [0]{}%
\providecommand \bibitemStop [0]{}%
\providecommand \bibitemNoStop [0]{.\EOS\space}%
\providecommand \EOS [0]{\spacefactor3000\relax}%
\providecommand \BibitemShut  [1]{\csname bibitem#1\endcsname}%
\let\auto@bib@innerbib\@empty
\bibitem [{\citenamefont {Wilson}\ and\ \citenamefont
  {Kogut}(1974)}]{Wilson197475}%
  \BibitemOpen
  \bibfield  {author} {\bibinfo {author} {\bibfnamefont {K.~G.}\ \bibnamefont
  {Wilson}}\ and\ \bibinfo {author} {\bibfnamefont {J.}~\bibnamefont {Kogut}},\
  }\href@noop {} {\bibfield  {journal} {\bibinfo  {journal} {Phys. Rep.}\
  }\textbf {\bibinfo {volume} {12}},\ \bibinfo {pages} {75} (\bibinfo {year}
  {1974})}\BibitemShut {NoStop}%
\bibitem [{\citenamefont {Stanley}(1988)}]{Stanley}%
  \BibitemOpen
  \bibfield  {author} {\bibinfo {author} {\bibfnamefont {H.~E.}\ \bibnamefont
  {Stanley}},\ }\href@noop {} {\emph {\bibinfo {title} {Introduction to Phase
  Transitions and Critical Phenomena}}}\ (\bibinfo  {publisher} {Oxford
  University Pres},\ \bibinfo {year} {1988})\BibitemShut {NoStop}%
\bibitem [{\citenamefont {Vinod}(1997)}]{G.Vinod}%
  \BibitemOpen
  \bibfield  {author} {\bibinfo {author} {\bibfnamefont {G.}~\bibnamefont
  {Vinod}},\ }\href@noop {} {\emph {\bibinfo {title} {PhD Thesis}}}\ (\bibinfo
  {publisher} {Cochin University of Science and Technology, Department of
  Physics},\ \bibinfo {year} {1997})\BibitemShut {NoStop}%
\bibitem [{\citenamefont {Sobhani}\ and\ \citenamefont
  {Hassanabadi}(2017)}]{Adv.High.EnergyPhys.20179530874}%
  \BibitemOpen
  \bibfield  {author} {\bibinfo {author} {\bibfnamefont {C.~W.~S.}\
  \bibnamefont {Sobhani}, \bibfnamefont {H.}}\ and\ \bibinfo {author}
  {\bibfnamefont {H.}~\bibnamefont {Hassanabadi}},\ }\href@noop {} {\bibfield
  {journal} {\bibinfo  {journal} {Adv. High Energy Phys.}\ ,\ \bibinfo {pages}
  {9530874}} (\bibinfo {year} {2017})}\BibitemShut {NoStop}%
\bibitem [{\citenamefont {Chung}\ and\ \citenamefont
  {Hassanabadi}(2017{\natexlab{a}})}]{Eur.Phys.J.Plus1322017398}%
  \BibitemOpen
  \bibfield  {author} {\bibinfo {author} {\bibfnamefont {S.~H.}\ \bibnamefont
  {Chung}, \bibfnamefont {W.~S.}}\ and\ \bibinfo {author} {\bibfnamefont
  {H.}~\bibnamefont {Hassanabadi}},\ }\href@noop {} {\bibfield  {journal}
  {\bibinfo  {journal} {Eur. Phys. J. Plus}\ }\textbf {\bibinfo {volume}
  {132}},\ \bibinfo {pages} {398} (\bibinfo {year}
  {2017}{\natexlab{a}})}\BibitemShut {NoStop}%
\bibitem [{\citenamefont {Chung}\ and\ \citenamefont
  {Hassanabadi}(2017{\natexlab{b}})}]{Int.J.Theor.Phys.5620171746}%
  \BibitemOpen
  \bibfield  {author} {\bibinfo {author} {\bibfnamefont {W.~S.}\ \bibnamefont
  {Chung}}\ and\ \bibinfo {author} {\bibfnamefont {H.}~\bibnamefont
  {Hassanabadi}},\ }\href@noop {} {\bibfield  {journal} {\bibinfo  {journal}
  {Int. J. Theor. Phys.}\ }\textbf {\bibinfo {volume} {56}},\ \bibinfo {pages}
  {1746} (\bibinfo {year} {2017}{\natexlab{b}})}\BibitemShut {NoStop}%
\bibitem [{\citenamefont {Du}\ and\ \citenamefont
  {Zhou}(2016)}]{EPL11320162000}%
  \BibitemOpen
  \bibfield  {author} {\bibinfo {author} {\bibfnamefont {X.~K.}\ \bibnamefont
  {Du}, \bibfnamefont {G.~J.}}\ and\ \bibinfo {author} {\bibfnamefont
  {C.}~\bibnamefont {Zhou}},\ }\href@noop {} {\bibfield  {journal} {\bibinfo
  {journal} {EPL}\ }\textbf {\bibinfo {volume} {113}},\ \bibinfo {pages}
  {20002} (\bibinfo {year} {2016})}\BibitemShut {NoStop}%
\bibitem [{\citenamefont {Dil}(2017)}]{Phys.DarkUniv.1620171}%
  \BibitemOpen
  \bibfield  {author} {\bibinfo {author} {\bibfnamefont {E.}~\bibnamefont
  {Dil}},\ }\href@noop {} {\bibfield  {journal} {\bibinfo  {journal} {Phys.
  Dark Univ.}\ }\textbf {\bibinfo {volume} {16}},\ \bibinfo {pages} {1}
  (\bibinfo {year} {2017})}\BibitemShut {NoStop}%
\bibitem [{\citenamefont {Boumali}\ and\ \citenamefont
  {Hassanabadi}(2017)}]{Adv.HighEnergyPhys.20179371391}%
  \BibitemOpen
  \bibfield  {author} {\bibinfo {author} {\bibfnamefont {A.}~\bibnamefont
  {Boumali}}\ and\ \bibinfo {author} {\bibfnamefont {H.}~\bibnamefont
  {Hassanabadi}},\ }\href@noop {} {\bibfield  {journal} {\bibinfo  {journal}
  {Adv. High Energy Phys.}\ ,\ \bibinfo {pages} {9371391}} (\bibinfo {year}
  {2017})}\BibitemShut {NoStop}%
\bibitem [{\citenamefont {Watanabe}(2016)}]{JHEP1220160630}%
  \BibitemOpen
  \bibfield  {author} {\bibinfo {author} {\bibfnamefont {N.}~\bibnamefont
  {Watanabe}},\ }\href@noop {} {\bibfield  {journal} {\bibinfo  {journal}
  {JHEP}\ }\textbf {\bibinfo {volume} {12}},\ \bibinfo {pages} {063} (\bibinfo
  {year} {2016})}\BibitemShut {NoStop}%
\bibitem [{\citenamefont {Dey}(2015)}]{Phys.Rev.D912015044024}%
  \BibitemOpen
  \bibfield  {author} {\bibinfo {author} {\bibfnamefont {S.}~\bibnamefont
  {Dey}},\ }\href@noop {} {\bibfield  {journal} {\bibinfo  {journal} {Phys.
  Rev. D}\ }\textbf {\bibinfo {volume} {91}},\ \bibinfo {pages} {044024}
  (\bibinfo {year} {2015})}\BibitemShut {NoStop}%
\bibitem [{\citenamefont {Zinn-Justin}(2002)}]{ZinnJustin}%
  \BibitemOpen
  \bibfield  {author} {\bibinfo {author} {\bibfnamefont {J.}~\bibnamefont
  {Zinn-Justin}},\ }\href@noop {} {\emph {\bibinfo {title} {Quantum Field
  Theory and Critical Phenomena}}}\ (\bibinfo  {publisher} {International
  Series of Monographs on Physics, Oxford University Press},\ \bibinfo {year}
  {2002})\BibitemShut {NoStop}%
\bibitem [{\citenamefont {Brezin}\ \emph {et~al.}(1976)\citenamefont {Brezin},
  \citenamefont {\mbox{Le Guillou}},\ and\ \citenamefont
  {Zinn-Justin}}]{BrezinLeGuillouZinnJustin}%
  \BibitemOpen
  \bibfield  {author} {\bibinfo {author} {\bibfnamefont {E.}~\bibnamefont
  {Brezin}}, \bibinfo {author} {\bibfnamefont {J.~C.}\ \bibnamefont {\mbox{Le
  Guillou}}}, \ and\ \bibinfo {author} {\bibfnamefont {J.}~\bibnamefont
  {Zinn-Justin}},\ }\href@noop {} {\emph {\bibinfo {title} {Phase Transitions
  and Critical Phenomena}}}\ (\bibinfo  {publisher} {Academic Press, London,
  edited by C. Domb and M. S. A. Green, Vol. 6, p. 125},\ \bibinfo {year}
  {1976})\BibitemShut {NoStop}%
\bibitem [{\citenamefont {Amit}\ and\ \citenamefont
  {Mart\'in-Mayor}(2005)}]{Amit}%
  \BibitemOpen
  \bibfield  {author} {\bibinfo {author} {\bibfnamefont {D.~J.}\ \bibnamefont
  {Amit}}\ and\ \bibinfo {author} {\bibfnamefont {V.}~\bibnamefont
  {Mart\'in-Mayor}},\ }\href@noop {} {\emph {\bibinfo {title} {Field Theory,
  The Renormalization Group and Critical Phenomena}}}\ (\bibinfo  {publisher}
  {World Scientific Pub Co Inc},\ \bibinfo {year} {2005})\BibitemShut {NoStop}%
\bibitem [{\citenamefont {Brezin}\ and\ \citenamefont
  {Zinn-Justin}(1973)}]{Phys.Rev.D841973434}%
  \BibitemOpen
  \bibfield  {author} {\bibinfo {author} {\bibfnamefont {L.~G. J.~C.}\
  \bibnamefont {Brezin}, \bibfnamefont {E.}}\ and\ \bibinfo {author}
  {\bibfnamefont {J.}~\bibnamefont {Zinn-Justin}},\ }\href@noop {} {\bibfield
  {journal} {\bibinfo  {journal} {Phys. Rev. D}\ }\textbf {\bibinfo {volume}
  {84}},\ \bibinfo {pages} {434} (\bibinfo {year} {1973})}\BibitemShut
  {NoStop}%
\bibitem [{\citenamefont {Carvalho}\ and\ \citenamefont
  {Leite}(2009)}]{Carvalho2009178}%
  \BibitemOpen
  \bibfield  {author} {\bibinfo {author} {\bibfnamefont {P.~R.~S.}\
  \bibnamefont {Carvalho}}\ and\ \bibinfo {author} {\bibfnamefont {M.~M.}\
  \bibnamefont {Leite}},\ }\href@noop {} {\bibfield  {journal} {\bibinfo
  {journal} {Ann. Phys.}\ }\textbf {\bibinfo {volume} {324}},\ \bibinfo {pages}
  {178} (\bibinfo {year} {2009})}\BibitemShut {NoStop}%
\bibitem [{\citenamefont {Carvalho}\ and\ \citenamefont
  {Leite}(2010)}]{Carvalho2010151}%
  \BibitemOpen
  \bibfield  {author} {\bibinfo {author} {\bibfnamefont {P.~R.~S.}\
  \bibnamefont {Carvalho}}\ and\ \bibinfo {author} {\bibfnamefont {M.~M.}\
  \bibnamefont {Leite}},\ }\href@noop {} {\bibfield  {journal} {\bibinfo
  {journal} {Ann. Phys.}\ }\textbf {\bibinfo {volume} {325}},\ \bibinfo {pages}
  {151} (\bibinfo {year} {2010})}\BibitemShut {NoStop}%
\bibitem [{\citenamefont {Ramond}(2001)}]{Ramond}%
  \BibitemOpen
  \bibfield  {author} {\bibinfo {author} {\bibfnamefont {P.}~\bibnamefont
  {Ramond}},\ }\href@noop {} {\emph {\bibinfo {title} {Field Theory: A Modern
  Primer}}}\ (\bibinfo  {publisher} {Frontiers in Physics Series, Vol. 74,
  Revised Printing, Westview Press},\ \bibinfo {year} {2001})\BibitemShut
  {NoStop}%
\bibitem [{\citenamefont {Carvalho}\ and\ \citenamefont
  {Leite}(2013)}]{J.Math.Phys.542013093301}%
  \BibitemOpen
  \bibfield  {author} {\bibinfo {author} {\bibfnamefont {P.}~\bibnamefont
  {Carvalho}}\ and\ \bibinfo {author} {\bibfnamefont {M.}~\bibnamefont
  {Leite}},\ }\href@noop {} {\bibfield  {journal} {\bibinfo  {journal} {J.
  Math. Phys.}\ }\textbf {\bibinfo {volume} {54}},\ \bibinfo {pages} {093301}
  (\bibinfo {year} {2013})}\BibitemShut {NoStop}%
\bibitem [{\citenamefont {Bogoliubov}\ and\ \citenamefont
  {Parasyuk}(1957)}]{BogoliubovParasyuk}%
  \BibitemOpen
  \bibfield  {author} {\bibinfo {author} {\bibfnamefont {N.~N.}\ \bibnamefont
  {Bogoliubov}}\ and\ \bibinfo {author} {\bibfnamefont {O.~S.}\ \bibnamefont
  {Parasyuk}},\ }\href@noop {} {\bibfield  {journal} {\bibinfo  {journal} {Acta
  Math.}\ }\textbf {\bibinfo {volume} {97}},\ \bibinfo {pages} {227} (\bibinfo
  {year} {1957})}\BibitemShut {NoStop}%
\bibitem [{\citenamefont {Hepp}(1966)}]{Hepp}%
  \BibitemOpen
  \bibfield  {author} {\bibinfo {author} {\bibfnamefont {K.}~\bibnamefont
  {Hepp}},\ }\href@noop {} {\bibfield  {journal} {\bibinfo  {journal} {Commun.
  Math. Phys.}\ }\textbf {\bibinfo {volume} {2}},\ \bibinfo {pages} {301}
  (\bibinfo {year} {1966})}\BibitemShut {NoStop}%
\bibitem [{\citenamefont {Zimmermann}(1969)}]{Zimmermann}%
  \BibitemOpen
  \bibfield  {author} {\bibinfo {author} {\bibfnamefont {W.}~\bibnamefont
  {Zimmermann}},\ }\href@noop {} {\bibfield  {journal} {\bibinfo  {journal}
  {Commun. Math. Phys.}\ }\textbf {\bibinfo {volume} {15}},\ \bibinfo {pages}
  {208} (\bibinfo {year} {1969})}\BibitemShut {NoStop}%
\bibitem [{\citenamefont {Kleinert}\ and\ \citenamefont
  {Schulte-Frohlinde}(2001)}]{Kleinert}%
  \BibitemOpen
  \bibfield  {author} {\bibinfo {author} {\bibfnamefont {H.}~\bibnamefont
  {Kleinert}}\ and\ \bibinfo {author} {\bibfnamefont {V.}~\bibnamefont
  {Schulte-Frohlinde}},\ }\href@noop {} {\emph {\bibinfo {title} {Critical
  Properties of $\phi^{4}$ Theories}}}\ (\bibinfo  {publisher} {World
  Scientific Pub Co Inc},\ \bibinfo {year} {2001})\BibitemShut {NoStop}%
\bibitem [{\citenamefont {Kosteleck\'{y}}\ and\ \citenamefont
  {Tasson}(2011)}]{Phys.Rev.D832011016013}%
  \BibitemOpen
  \bibfield  {author} {\bibinfo {author} {\bibfnamefont {V.~A.}\ \bibnamefont
  {Kosteleck\'{y}}}\ and\ \bibinfo {author} {\bibfnamefont {J.~D.}\
  \bibnamefont {Tasson}},\ }\href@noop {} {\bibfield  {journal} {\bibinfo
  {journal} {Phys. Rev. D.}\ }\textbf {\bibinfo {volume} {83}},\ \bibinfo
  {pages} {016013} (\bibinfo {year} {2011})}\BibitemShut {NoStop}%
\bibitem [{\citenamefont {Ferrero}\ and\ \citenamefont
  {Altschul}(2011)}]{PhysRevD.84.065030}%
  \BibitemOpen
  \bibfield  {author} {\bibinfo {author} {\bibfnamefont {A.}~\bibnamefont
  {Ferrero}}\ and\ \bibinfo {author} {\bibfnamefont {B.}~\bibnamefont
  {Altschul}},\ }\href@noop {} {\bibfield  {journal} {\bibinfo  {journal}
  {Phys. Rev. D}\ }\textbf {\bibinfo {volume} {84}},\ \bibinfo {pages} {065030}
  (\bibinfo {year} {2011})}\BibitemShut {NoStop}%
\bibitem [{\citenamefont {Carvalho}(2013)}]{Carvalho2013850}%
  \BibitemOpen
  \bibfield  {author} {\bibinfo {author} {\bibfnamefont {P.~R.~S.}\
  \bibnamefont {Carvalho}},\ }\href@noop {} {\bibfield  {journal} {\bibinfo
  {journal} {Phys. Lett. B}\ }\textbf {\bibinfo {volume} {726}},\ \bibinfo
  {pages} {850} (\bibinfo {year} {2013})}\BibitemShut {NoStop}%
\bibitem [{\citenamefont {Carvalho}(2014)}]{Carvalho2014320}%
  \BibitemOpen
  \bibfield  {author} {\bibinfo {author} {\bibfnamefont {P.~R.~S.}\
  \bibnamefont {Carvalho}},\ }\href@noop {} {\bibfield  {journal} {\bibinfo
  {journal} {Phys. Lett. B}\ }\textbf {\bibinfo {volume} {730}},\ \bibinfo
  {pages} {320} (\bibinfo {year} {2014})}\BibitemShut {NoStop}%
\bibitem [{\citenamefont {Vieira}\ and\ \citenamefont
  {Carvalho}(2014)}]{EurophysLett.108.21001}%
  \BibitemOpen
  \bibfield  {author} {\bibinfo {author} {\bibfnamefont {W.~C.}\ \bibnamefont
  {Vieira}}\ and\ \bibinfo {author} {\bibfnamefont {P.~R.~S.}\ \bibnamefont
  {Carvalho}},\ }\href@noop {} {\bibfield  {journal} {\bibinfo  {journal}
  {Europhys. Lett.}\ }\textbf {\bibinfo {volume} {108}},\ \bibinfo {pages}
  {21001} (\bibinfo {year} {2014})}\BibitemShut {NoStop}%
\bibitem [{\citenamefont {Carvalho}(2016)}]{Int.J.Mod.Phys.B.30.1550259}%
  \BibitemOpen
  \bibfield  {author} {\bibinfo {author} {\bibfnamefont {P.~R.~S.}\
  \bibnamefont {Carvalho}},\ }\href@noop {} {\bibfield  {journal} {\bibinfo
  {journal} {Int. J. Mod. Phys. B}\ }\textbf {\bibinfo {volume} {30}},\
  \bibinfo {pages} {1550259} (\bibinfo {year} {2016})}\BibitemShut {NoStop}%
\bibitem [{\citenamefont {Vieira}\ and\ \citenamefont
  {de~Carvalho}(2016)}]{Int.J.Geom.MethodsMod.Phys.13.1650049}%
  \BibitemOpen
  \bibfield  {author} {\bibinfo {author} {\bibfnamefont {W.~d.~C.}\
  \bibnamefont {Vieira}}\ and\ \bibinfo {author} {\bibfnamefont {P.~R.~S.}\
  \bibnamefont {de~Carvalho}},\ }\href@noop {} {\bibfield  {journal} {\bibinfo
  {journal} {Int. J. Geom. Methods Mod. Phys.}\ }\textbf {\bibinfo {volume}
  {13}},\ \bibinfo {pages} {1650049} (\bibinfo {year} {2016})}\BibitemShut
  {NoStop}%
\bibitem [{\citenamefont {Ryder}(1996)}]{Ryder}%
  \BibitemOpen
  \bibfield  {author} {\bibinfo {author} {\bibfnamefont {L.~H.}\ \bibnamefont
  {Ryder}},\ }\href@noop {} {\emph {\bibinfo {title} {Quantum Field Theory}}}\
  (\bibinfo  {publisher} {Cambridge University Press, 2nd Editio},\ \bibinfo
  {year} {1996})\BibitemShut {NoStop}%
\end{thebibliography}%

\end{document}